\documentclass[prd,showpacs,preprintnumbers,amsmath,amssymb,11pt]{revtex4}
\usepackage{graphics}
\usepackage{epsfig}
\usepackage{subfigure}
\usepackage{dcolumn}% Align table columns on decimal point
\usepackage{bm}% bold math
\usepackage{color}

\begin{document}

\title{Production of the $X_b$ in $\Upsilon(5S, 6S)\to \gamma X_b$ radiative decays}

\author{Qi Wu, Gang Li\footnote{gli@mail.qfnu.edu.cn}, Fenglan Shao, Qianwen Wang, Ruiqin Wang, Yawei Zhang and Ying Zheng
}

\affiliation{$^1$Department of Physics, Qufu Normal University, Qufu 273165, China }

\begin{abstract}
In this work, we investigate the production of  $X_b$ in the process $\Upsilon(5S,6S)\to \gamma X_b$, where $X_b$ is assumed to be the counterpart of $X(3872)$ in the bottomonium sector as a $B {\bar B}^*$ molecular state. We use the effective Lagrangian based on the heavy quark symmetry to explore the rescattering mechanism and calculate their production ratios. Our results have shown that the production ratios for the $\Upsilon(5S,6S) \to \gamma X_b$ are orders of $10^{-5}$ with reasonable cutoff parameter range $\alpha \simeq 2\sim 3$. The sizeable production ratios may be accessible at the future experiments like forthcoming BelleII, which will provide important clues to the inner structures of the exotic state $X_b$.
\end{abstract}

\date{\today}


%14.40.Rt Exotic mesons

%13.75.Lb Meson-meson interactions

%13.20.Gd Decays of J/\psi, and other quarkonia

%14.40.Pq Heavy quarkonia

%14.40.Lb Charmed mesons

\maketitle

\section{Introduction}
\label{sec:introduction}

In the past decades, many so called XYZ have been observed by the Belle, BaBar, CDF, D0, CMS, LHCb, and BESIII collaborations~\cite{Agashe:2014kda}. Some of them cannot fit into the conventional heavy quarkonium  in  the quark model~\cite{Brambilla:2010cs,Godfrey:2008nc,Drenska:2010kg,Bodwin:2013nua}. Up to now, many studies on the production and decay of these XYZ states have been carried out in order to understand its nature (for a recent review, see Refs.~\cite{Chen:2013wva,Liu:2013waa,Chen:2016qju}).

In 2003, the Belle Collaboration discovered an exotic candidate $X(3872)$  in the process $B^+\to K^++ J/\psi \pi^+\pi^-$~\cite{Choi:2003ue} which was subsequently confirmed by the BaBar Collaboration~\cite{Aubert:2004ns} in the same channel. It was also discovered
in proton-proton/antiproton collisions at the Tevatron~\cite{Abazov:2004kp,Aaltonen:2009vj} and LHC~\cite{Chatrchyan:2013cld,Aaij:2013zoa}. The $X(3872)$ is a particularly intriguing state because on the one hand its total width $\Gamma<1.2$ MeV~\cite{Agashe:2014kda} is tiny compared to typical hadronic widths; on the other
hand the closeness of its mass to the $D^0\overline D^{*0}$ threshold ($M_{X(3872)}-M_{D^0}-M_{D^{*0} }=(-0.12\pm0.24)$~MeV) and its prominent decays to $D^0\overline D^{*0}$~\cite{Agashe:2014kda} suggest that it may be an meson-meson molecular state~\cite{Tornqvist:2004qy,Hanhart:2007yq}.

Many theoretical works have been carried out in order to understand the nature of $X(3872)$ since the first observation of $X(3872)$.  It is also natural to look for the counterpart with $J^{PC}=1^{++}$ (denoted as $X_b$ hereafter) in the bottom sector. These two states are related by heavy quark symmetry which should have some universal properties.  The search for
$X_b$ may provide us important information on the discrimination of a compact multiquark configuration and a loosely bound hadronic molecule configuration.  Since the mass of $X_b$ may be very heavy and its $J^{PC}$ is $1^{++}$, it is less likely  for a direct discovery at the current electron-positron collision facilities, though the Super KEKB may provide an opportunity in $\Upsilon(5S,6S)$ radiative decays~\cite{Aushev:2010bq}.  In Ref.~\cite{He:2014sqj}, a search for $X_b$ in the $\omega \Upsilon(1S)$ final states has been presented and no significant signal
is observed for such a state.

The production of $X_b$ at the LHC and the Tevatron~\cite{Guo:2014sca,Guo:2013ufa} and other exotic states at hadron colliders~\cite{Bignamini:2009sk,Artoisenet:2009wk,Artoisenet:2010uu,Esposito:2013ada,Ali:2011qi,Ali:2013xba} have been extensively investigated. In the bottomonium system, the  isospin is almost perfectly conserved, which may explain the escape of $X_b$ in the recent CMS search~\cite{Chatrchyan:2013mea}. As a result, the radiative decays and isospin conserving decays will be of high priority in searching for $X_b$~\cite{Li:2014uia,Li:2015uwa,Karliner:2014sja}.
In Ref.~\cite{Li:2014uia}, we have studied the
radiative decays of $X_b \to \gamma \Upsilon(nS)$ ($n=1, 2, 3$), with $X_b$
being a candidate for the $B{\bar B}^*$ molecular state, and
found that the partial widths into $\gamma X_b$ are about $1$ keV. In Ref.~\cite{Li:2015uwa}, we studied the rescattering mechanism of the isospin conserving decays $X_b\to \Upsilon(1S)\omega$, and our results show that the partial width for the $X_b\to \Upsilon(1S)\omega$ is about tens of keVs.

In this work, we will further investigate the $X_b$ production in $\Upsilon(5S,6S) \to \gamma X_b$ with $X_b$ being a $B{\bar B}^*$ molecule candidate.  To investigate this process, we calculate the intermediate meson loop (IML) contributions. As well know, IML transitions have been one of the important nonperturbative transition mechanisms been noticed for a long time~\cite{Lipkin:1986bi,Lipkin:1988tg,Moxhay:1988ri}.
Recently, this mechanism has been used to study the production and decays of ordinary and  exotic states~\cite{Guo:2009wr,Guo:2010ak,Wang:2013cya,Liu:2013vfa,Guo:2013zbw,Wang:2013hga,Cleven:2013sq,Chen:2011pv,Li:2012as,Li:2013yla,Li:2014gxa,Li:2014pfa,Li:2013jma,Li:2013xia,Li:2013zcr,Li:2011ssa,Voloshin:2013ez,Voloshin:2011qa,Bondar:2011ev,oai:arXiv.org:1002.2712,Chen:2011pu,Chen:2012yr,Chen:2013coa,Chen:2013bha,Wang:2012mf,Yuan-Jiang:2010cna,Li:2010zf} and B decays~\cite{Du:1998ss,Chen:2000ih,Liu:2007qs,Lu:2005mx,Colangelo:2003sa,Liu:2008tv,Cheng:2004ru,Colangelo:2002mj}, and a global agreement with experimental data were obtained. Thus this approach may be suitable for the process $\Upsilon(5S,6S) \to \gamma X_b$.

The paper is organized as follows. In Sec.~\ref{sec:formula}, the
effective Lagrangians for our calculation.  Then in Sec.~\ref{sec:results}, we present our numerical results. Finally we give the summary in
Sec.~\ref{sec:summary}.

%%%%%%%%%%%%%%%%%%%%%
\section{Effective Lagrangians}
\label{sec:formula}
%%%%%%%%%%%%%%%%%%%%%

%%%%%%%%%%%%%%%%%%%%%%%%%%%%%%
%%%%%%%%%%%%%%%%%%%%%%%%%%%%%%
\begin{figure}[hbt]
\begin{center}
\includegraphics[width=0.8\textwidth]{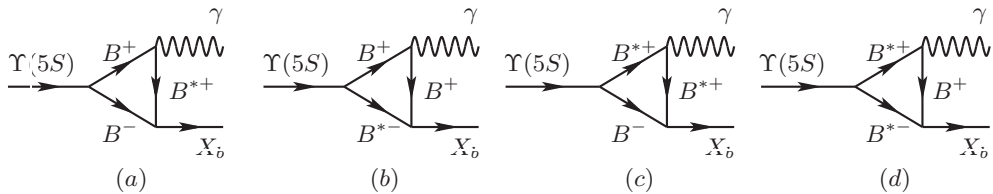}
\vglue-0mm\caption{Feynman  diagrams for $X_b$ production in $\Upsilon(5S) \to \gamma X_b  $ under the $B{\bar B}^*$ meson loop effects.} \label{fig:loops}
\end{center}
\end{figure}
%%%%%%%%%%%%%%%%%%%%%%%%%%%%%%
%%%%%%%%%%%%%%%%%%%%%%%%%%%%%%

Based on the  heavy quark symmetry, we can write out the relevant
effective Lagrangian for the $\Upsilon(5S)$~\cite{Colangelo:2003sa,Casalbuoni:1996pg}
\begin{eqnarray}
%%Upsilon B(*)B(*)
\mathcal{L}_{\Upsilon(5S) B^{(*)} B^{(*)}} &=&
ig_{\Upsilon BB} \Upsilon_{\mu} (\partial^\mu B \bar{B}- B
\partial^\mu \bar{B})-g_{\Upsilon B^* B} \varepsilon_{\mu \nu
\alpha \beta}
\partial^{\mu} \Upsilon^{\nu} (\partial^{\alpha} B^{*\beta} \bar{B}
 + B \partial^{\alpha}
\bar{B}^{*\beta})\nonumber\\
&&-ig_{\Upsilon B^* B^*} \big\{
\Upsilon^\mu (\partial_{\mu} B^{* \nu} \bar{B}^*_{\nu}
-B^{* \nu} \partial_{\mu}
\bar{B}^*_{\nu})+ (\partial_{\mu} \Upsilon_{\nu} B^{* \nu} -\Upsilon_{\nu}
\partial_{\mu} B^{* \nu}) \bar{B}^{* \mu}  \nonumber\\
&& +
B^{* \mu}(\Upsilon^\nu \partial_{\mu} \bar{B}^*_{\nu} -
\partial_{\mu} \Upsilon^\nu \bar{B}^*_{\nu})\big\}, \label{eq:h1}
\end{eqnarray}
where
${{B}^{(*)}}=\left(B^{(*)+},B^{(*)0}\right)$ and
${\bar B^{(*)T}}=\left(B^{(*)-},\bar{B}^{(*)0}\right)$ correspond to the
bottom meson isodoublets. $\epsilon_{\mu\nu\alpha\beta}$ is the anti-symmetric Levi-Civita tensor and $\epsilon_{0123}= +1$. Since $\Upsilon(5S)$ is above the threshold of $B^{(*)} {\bar B}^{(*)}$, the coupling constants between $\Upsilon(5S)$ and $B^{(*)} {\bar B}^{(*)}$ can be determined via experimental data for $\Upsilon(5S) \to B^{(*)} {\bar B}^{(*)}$~\cite{Agashe:2014kda}. The experimental branching ratios and the corresponding coupling constants are listed in Table~\ref{tab:coupling constants}. Since there is no experimental information on $\Upsilon(6S) \to B^{(*)} {\bar B}^{(*)}$~\cite{Agashe:2014kda}, we choose the coupling constants between $\Upsilon(6S)$ and $B^{(*)} {\bar B}^{(*)}$ the same values as that of $\Upsilon(5S)$.

%%%%%%%%%%%%%%%%%%%%%%%%%%%%%%%%%%%%%%%
\begin{table}[htb]
\begin{center}
\caption{The coupling constants of $\Upsilon(5S)$ interacting with $B^{(*)}{\bar B}^{(*)}$. Here, we list the corresponding branching ratios of $\Upsilon(5S)\to B^{(*)}{\bar B}^{(*)}$. }\label{tab:coupling constants}
 \begin{tabular}{ccccccccc}
 \hline
 Final state  & $\mathcal{B}(\%)$ & Coupling & Final state & $\mathcal{B}(\%)$ & Coupling & Final state & $\mathcal{B}(\%)$ & Coupling\\ \hline
 $B {\bar B}$ & $5.5$  & $1.76$  & $B {\bar B}^*+c.c.$ & $13.7$  & $0.14$ GeV$^{-1}$  & $B^* {\bar B}^*$ & $38.1$  & $2.22$ \\ \hline
 $B_s {\bar B}_s$ & $0.5$  & $0.96$  & $B_s {\bar B}_s^*+c.c.$ & $1.35$  & $0.10$ GeV$^{-1}$  & $B_s^* {\bar B}_s^*$ & $17.6$  & $5.07$ \\ \hline
\end{tabular}
\end{center}
\end{table}
%%%%%%%%%%%%%%%%%%%%%%%%%%%%%%%%%%%%%%%%%%%%%%%%%%%%%%%%%
In order to calculate the process depicted in Fig.~\ref{fig:loops}, we also need the photonic coupling to the bottomed mesons. The magnetic coupling of the photon to heavy bottom  meson is described by the Lagrangian~\cite{Hu:2005gf,Amundson:1992yp}
\begin{eqnarray}
{\cal L}_\gamma = \frac {e\beta Q_{ab}} {2} F^{\mu\nu} {\rm Tr}[H_b^\dagger \sigma_{\mu\nu} H_a  ] + \frac {e Q^\prime} {2m_{Q}} F^{\mu\nu} {\rm Tr}[H_a^\dagger H_a \sigma_{\mu\nu}], \label{eq:coupling-photon}
\end{eqnarray}
with
\begin{eqnarray}
H&=&\left( \frac{1+ \rlap{/}{v} }{2} \right)
[\mathcal{B}^{*\mu}
\gamma_\mu -\mathcal{B}\gamma_5],
\end{eqnarray}
where $\beta$ is an unknown constant, $Q= {\rm diag}\{2/3, -1/3, -1/3\}$ is the light quark charge
matrix,  and $Q^\prime$ is the heavy quark
electric charge (in units of $e$).  $\beta\simeq 3.0$ GeV$^{-1}$ is determined in the nonrelativistic constituent quark model and has been adopted  in the study of radiative $D^*$ decays~\cite{Amundson:1992yp}. In the $b$ and $c$ systems, the $\beta$ value is the same due to heavy quark symmetry~\cite{Amundson:1992yp}. In Eq.~(\ref{eq:coupling-photon}), the first term is the magnetic
moment coupling of the light quarks, while the second one  is the
magnetic moment coupling of the heavy quark and hence is
suppressed by $1/m_Q$.

At last, assuming that  $X_b$ is an $S$-wave molecule with $J^{PC}=1^{++}$ given by the superposition of $B^0 {\bar B}^{*0}+c.c$ and $B^- {\bar B}^{*+}+c.c$ hadronic configurations as
\begin{eqnarray}
|X_b\rangle= \frac {1} {2} [  (|B^0{\bar B}^{*0}\rangle - |B^{*0} {\bar B}^0\rangle) +   (| B^+ B^{*-}\rangle - | B^- B^{*+}\rangle ) ].
\end{eqnarray}
As a result, we can parameterize the coupling of $X_b$ to the bottomed mesons in terms of the following Lagrangian
\begin{eqnarray}
{\cal L} = \frac {1} {2} X_{b\mu}^{\dagger} [x_1(B^{*0\mu} {\bar B}^0 - B^{0} {\bar B}^{*0\mu})+x_2(B^{*+\mu} B^- - B^+ B^{*-\mu})] + h.c.,
\end{eqnarray}
where $x_i$ denotes the  coupling constant. Since the $X_b$ is slightly below the $S$-wave $B{\bar B}^*$ threshold, the effective coupling of this
state is related to the probability of finding the $B{\bar B}^*$
component in the physical wave function of the bound
states and the binding energy, $\epsilon_{X_b}=m_B+m_{B^*}-m_{X_b}$~\cite{Weinberg:1965zz, Baru:2003qq,Guo:2013zbw}
\begin{eqnarray}\label{eq:coupling-Xb}
x_i^2 \equiv 16\pi (m_B+ m_{B^*})^2 c_i^2 \sqrt{\frac {2\epsilon_{X_b}}{\mu}} ,
\end{eqnarray}
where $c_i=1/{\sqrt 2}$, $\mu=m_B m_{B^*}/(m_B+m_{B^*})$ is the reduced mass. Here, we should also notice that the coupling constant $x_i$ in Eq.~(\ref{eq:coupling-Xb}) is based on the assumption that $X_b$ is a shallow bound state where the
potential binding the mesons is short-ranged.

Based on the relevant Lagrangians given above, the decay amplitudes in
Fig.~\ref{fig:loops} can be generally expressed as follows,
\begin{eqnarray}
M_{fi}=\int \frac {d^4 q_2} {(2\pi)^4} \sum_{B^* \ \mbox{pol.}}
\frac {T_1T_2T_3} {D_1 D_2 D_3}{\cal F}(m_2,q_2^2)
\end{eqnarray}
where $T_i$ and $D_i = q_i^2-m_i^2 \ (i=1,2,3)$ are the vertex
functions and the denominators of the intermediate meson
propagators. For example, in Fig.~\ref{fig:loops} (a), $T_i \
(i=1,2,3)$ are the vertex functions for the initial $\Upsilon(5S)$, final
$X_b$ and photon, respectively. $D_i \
(i=1,2,3)$  are the denominators for the intermediate $B^+$,
$B^{-}$ and $B^{*+}$ propagators, respectively.

Since the intermediate exchanged bottom mesons in the triangle diagram Fig.~\ref{fig:loops} are
off-shell, in order to
compensate this off-shell effects arising from
the intermediate exchanged particle and  also the non-local effects of the
vertex functions~\cite{Li:1996yn,Locher:1993cc,Li:1996cj}, we adopt the following form factors,
\begin{eqnarray}\label{ELA-form-factor}
{\cal F}(m_{2}, q_2^2) \equiv \left(\frac
{\Lambda^2-m_{2}^2} {\Lambda^2-q_2^2}\right)^n,
\end{eqnarray}
where $n=1,2$ corresponds monopole and dipole form factor, respectively. $\Lambda\equiv m_2+\alpha\Lambda_{\rm QCD}$ and the QCD energy
scale $\Lambda_{\rm QCD} = 220$ MeV. This form factor is supposed and many phenomenological studies have suggested $\alpha\simeq 2\sim 3$. These two form factors can help us explore  the dependence of our results on the form factor.

The explicit expression of transition amplitudes can be found in Appendix (A.2) in Ref.~\cite{Zhao:2013jza}, where radiative decays of charmonium are studied extensively based on effective Lagrangian approach.

%%%%%%%%%%%%%%%%%%%%%%%
\section{Numerical Results}
\label{sec:results}
%%%%%%%%%%%%%%%%%%%%%%%
%%%%%%%%%%%%%%%%%%%%%%%%%%%%%%%%%%%%%%%
\begin{table}[htb]
\begin{center}
\caption{Predicted branching ratios for $\Upsilon(5S) \to \gamma X_b $.  The  parameter in the form factor is chosen as $\alpha =2.0$, $2.5$, and $3.0$. The last column is the calculated branching ratios in NREFT approach. }\label{tab:results-5S}
 \begin{tabular}{cccccccc}
 \hline
Binding Energy & \multicolumn{3}{c}{Monopole form factor} & \multicolumn{3}{c}{Dipole form factor} & NREFT\\
          %\cline{2-7}
& $\alpha=2.0 $ & $\alpha=2.5 $ &$\alpha=3.0 $ & $\alpha=2.0 $ & $\alpha=2.5 $ &$\alpha=3.0 $ \\ \hline
$\epsilon_{X_b}=5$ MeV    & $2.02\times 10^{-5}$  & $2.06\times 10^{-5}$  & $2.08\times 10^{-5}$  & $1.90\times 10^{-5}$  & $1.99\times 10^{-5}$  & $2.04\times 10^{-5}$    & $1.52\times 10^{-6}$ \\ \hline
$\epsilon_{X_b}=10$ MeV   & $2.58\times 10^{-5}$  & $2.66\times 10^{-5}$  & $2.71\times 10^{-5}$  & $2.32\times 10^{-5}$  & $2.47\times 10^{-5}$  & $2.57\times 10^{-5}$    & $2.12\times 10^{-6}$ \\ \hline
$\epsilon_{X_b}=25$ MeV   & $3.24\times 10^{-5}$  & $3.42\times 10^{-5}$  & $3.54\times 10^{-5}$  & $2.61\times 10^{-5}$  & $2.90\times 10^{-5}$  & $3.09\times 10^{-5}$    & $3.88\times 10^{-6}$ \\ \hline
$\epsilon_{X_b}=50$ MeV   & $3.37\times 10^{-5}$  & $3.65\times 10^{-5}$  & $3.85\times 10^{-5}$  & $2.37\times 10^{-5}$  & $2.75\times 10^{-5}$  & $3.04\times 10^{-5}$    & $6.41\times 10^{-6}$\\ \hline
$\epsilon_{X_b}=100$ MeV  & $2.91\times 10^{-5}$  & $3.27\times 10^{-5}$  & $3.54\times 10^{-5}$  & $1.65\times 10^{-5}$  & $2.05\times 10^{-5}$  & $2.38\times 10^{-5}$    & $1.20\times 10^{-5}$\\ \hline
\end{tabular}
\end{center}
\end{table}
%%%%%%%%%%%%%%%%%%%%%%%%%%%%%%%%%%%%%%%%%%%%%%%%%%%%%%%%%

%%%%%%%%%%%%%%%%%%%%%%%%%%%%%%%%%%%%%%%
\begin{table}[htb]
\begin{center}
\caption{Predicted branching ratios for $\Upsilon(6S) \to \gamma X_b $.  The  parameter in the form factor is chosen as $\alpha =2.0$, $2.5$, and $3.0$. The last column is the calculated branching ratios in NREFT approach.}\label{tab:results-6S}
 \begin{tabular}{cccccccc}
 \hline
Binding Energy & \multicolumn{3}{c}{Monopole form factor} & \multicolumn{3}{c}{Dipole form factor}  & NREFT\\
          %\cline{2-7}
& $\alpha=2.0 $ & $\alpha=2.5 $ &$\alpha=3.0 $ & $\alpha=2.0 $ & $\alpha=2.5 $ &$\alpha=3.0 $ &\\ \hline
$\epsilon_{X_b}=5$ MeV    & $9.71\times 10^{-6}$  & $1.02\times 10^{-5}$  & $1.05\times 10^{-5}$  & $8.16\times 10^{-6}$  & $9.04\times 10^{-6}$  & $9.63\times 10^{-6}$  & $3.38\times 10^{-6}$ \\ \hline
$\epsilon_{X_b}=10$ MeV   & $1.25\times 10^{-5}$  & $1.33\times 10^{-5}$  & $1.38\times 10^{-5}$  & $9.97\times 10^{-6}$  & $1.13\times 10^{-5}$  & $1.22\times 10^{-5}$  & $4.89\times 10^{-6}$ \\ \hline
$\epsilon_{X_b}=25$ MeV   & $1.62\times 10^{-5}$  & $1.76\times 10^{-5}$  & $1.85\times 10^{-5}$  & $1.14\times 10^{-5}$  & $1.34\times 10^{-5}$  & $1.49\times 10^{-5}$  & $8.27\times 10^{-6}$ \\ \hline
$\epsilon_{X_b}=50$ MeV   & $1.76\times 10^{-5}$  & $1.96\times 10^{-5}$  & $2.12\times 10^{-5}$  & $1.08\times 10^{-5}$  & $1.32\times 10^{-5}$  & $1.52\times 10^{-5}$  & $1.30\times 10^{-5}$ \\ \hline
$\epsilon_{X_b}=100$ MeV  & $1.66\times 10^{-5}$  & $1.92\times 10^{-5}$  & $2.12\times 10^{-5}$  & $8.12\times 10^{-6}$  & $1.06\times 10^{-5}$  & $1.28\times 10^{-5}$  & $2.24\times 10^{-5}$ \\ \hline
\end{tabular}
\end{center}
\end{table}
%%%%%%%%%%%%%%%%%%%%%%%%%%%%%%%%%%%%%%%%%%%%%%%%%%%%%%%%%
%%%%%%%%%%%%%%%%%%%%%%%%%%%%%%
%%%%%%%%%%%%%%%%%%%%%%%%%%%%%%
\begin{figure}[hbt]
\begin{center}
\includegraphics[width=0.49\textwidth]{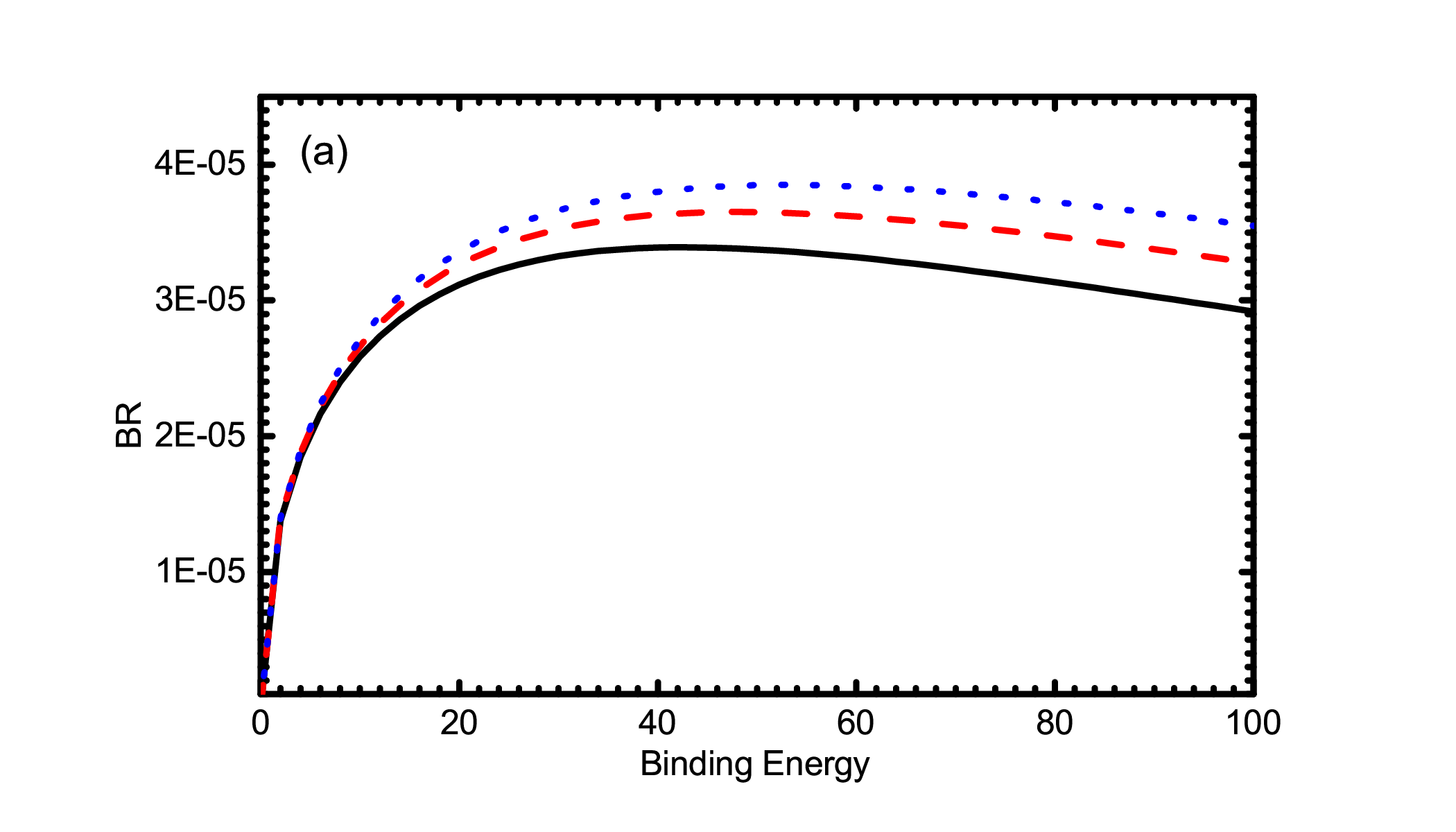}
\includegraphics[width=0.49\textwidth]{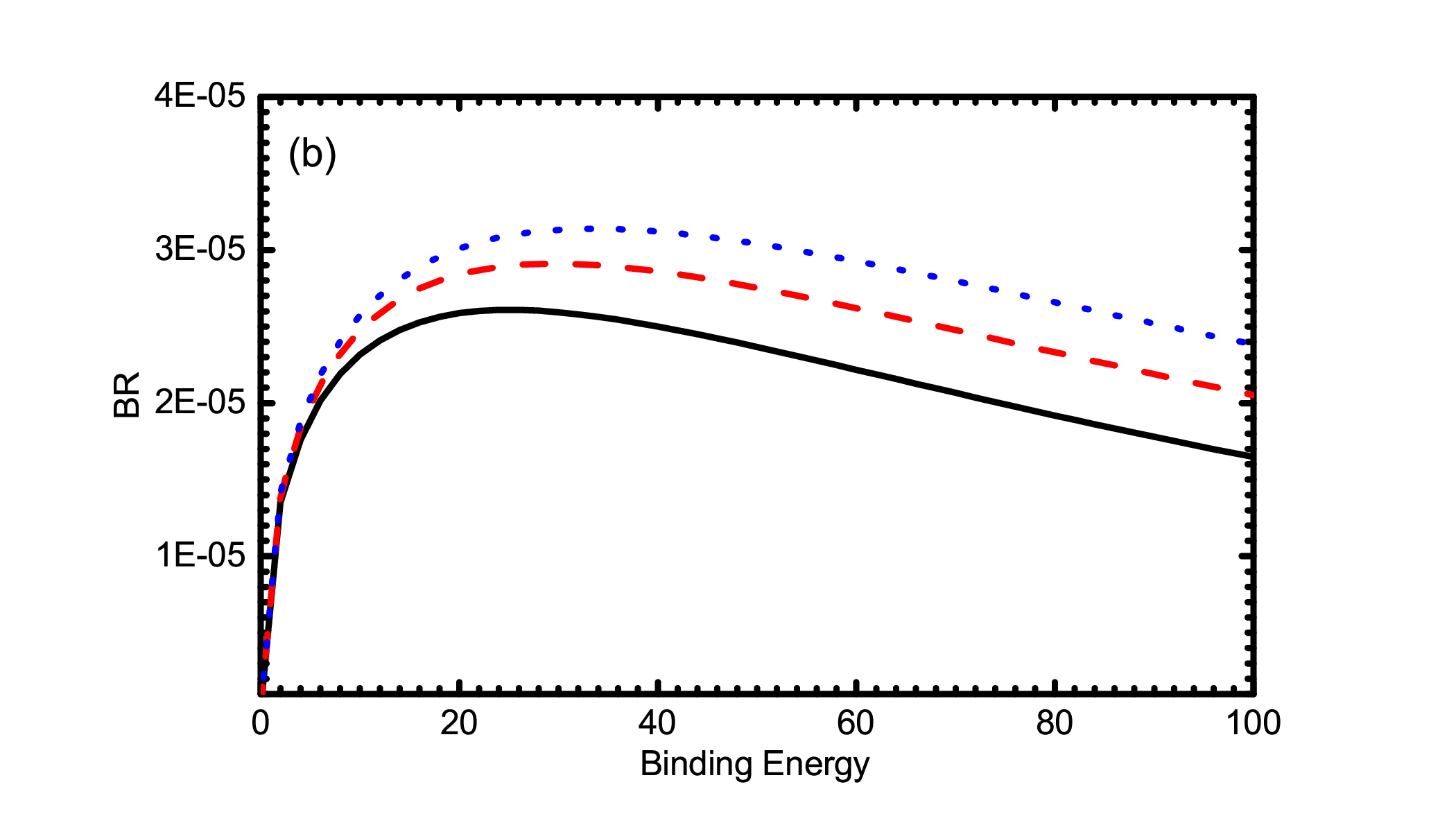}
\caption{(a). The dependence of the branching ratios of $\Upsilon(5S) \to \gamma X_b$ on the $\epsilon_{X_b}$ using monopole form factors with $\alpha =2.0$ (solid lines), $\alpha=2.5$ (dashed lines), and $\alpha=3.0$ (dotted lines), respectively. (b). The dependence of the branching ratios of $\Upsilon(5S) \to \gamma X_b $ on the $\epsilon_{X_b}$ using dipole form factors with $\alpha =2.0$ (solid lines), $\alpha=2.5$ (dashed lines), and $\alpha=3.0$ (dotted lines), respectively. The results with binding energy up to $100$ MeV might make the molecular state assumption inaccurate.} \label{fig:BrOnEXb}
\end{center}
\end{figure}
%%%%%%%%%%%%%%%%%%%%%%%%%%%%%%

%%%%%%%%%%%%%%%%%%%%%%%%%%%%%%
\begin{figure}[hbt]
\begin{center}
\includegraphics[width=0.49\textwidth]{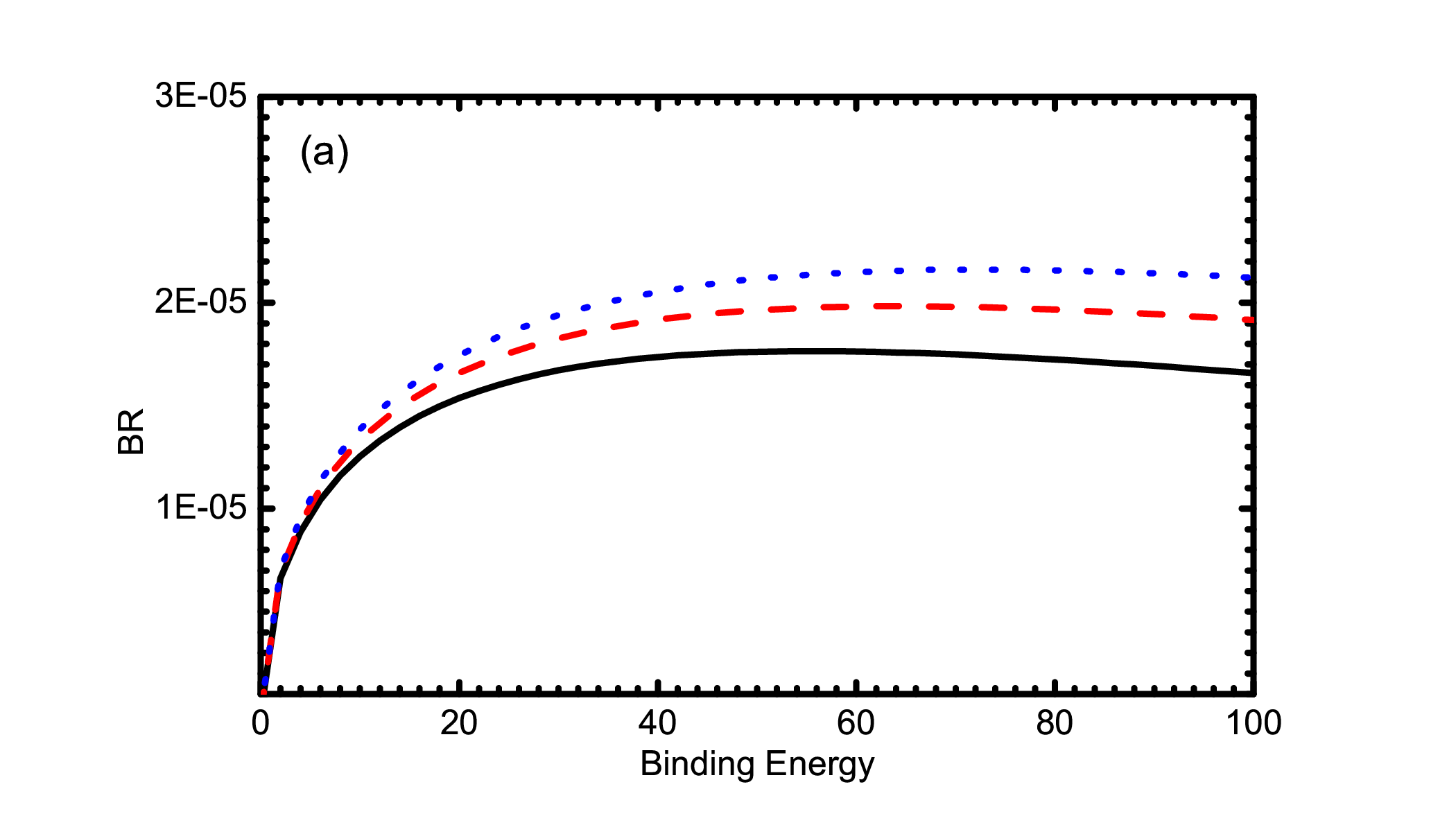}
\includegraphics[width=0.49\textwidth]{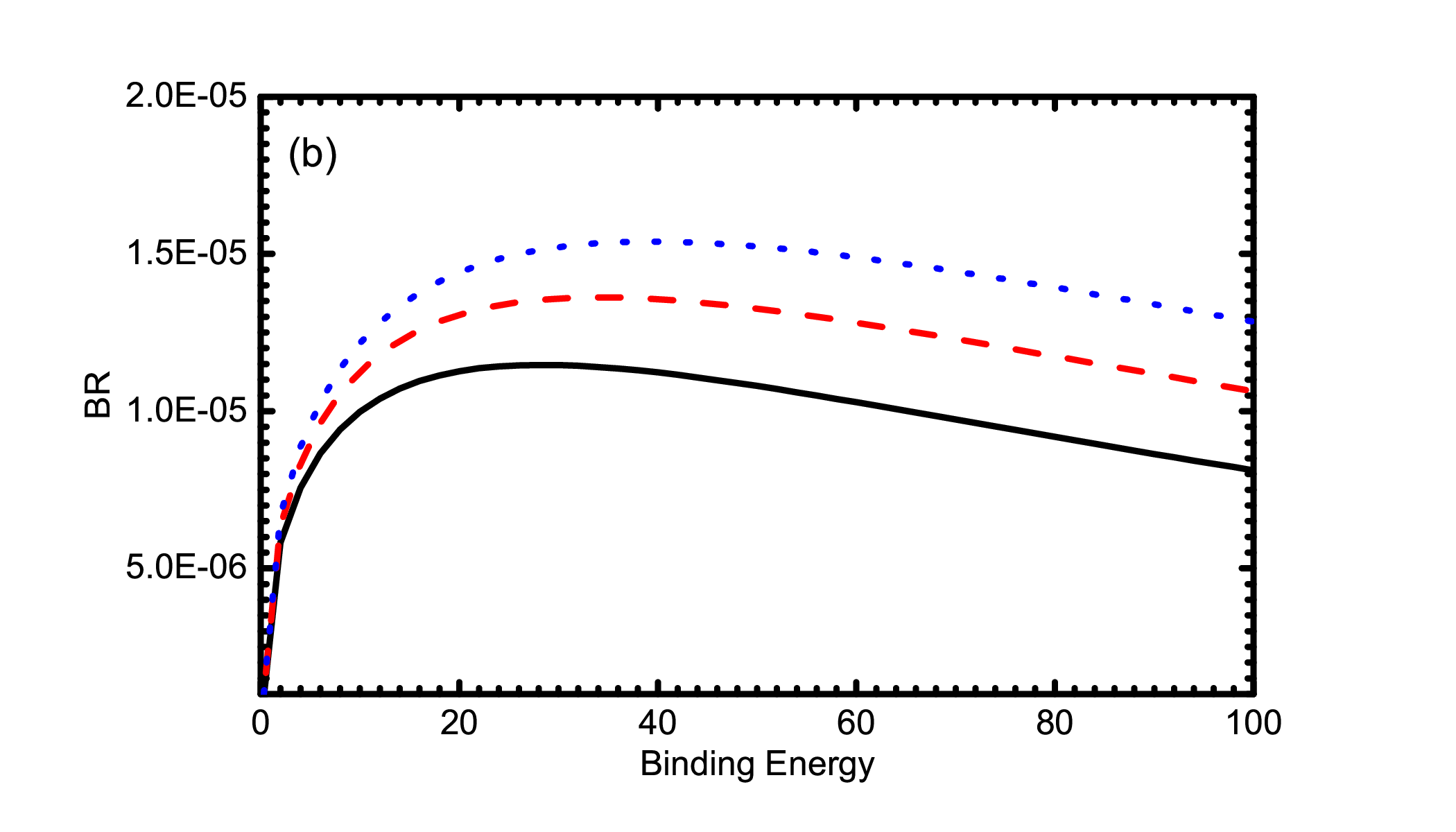}
\caption{(a). The dependence of the branching ratios of $\Upsilon(6S) \to \gamma X_b$ on the $\epsilon_{X_b}$ using monopole form factors with $\alpha =2.0$ (solid lines), $\alpha=2.5$ (dashed lines), and $\alpha=3.0$ (dotted lines), respectively. (b). The dependence of the branching ratios of $\Upsilon(6S) \to \gamma X_b $ on the $\epsilon_{X_b}$ using dipole form factors with $\alpha =2.0$ (solid lines), $\alpha=2.5$ (dashed lines), and $\alpha=3.0$ (dotted lines), respectively. The results with binding energy up to $100$ MeV might make the molecular state assumption inaccurate. } \label{fig:BrOnEXb_6S}
\end{center}
\end{figure}
%%%%%%%%%%%%%%%%%%%%%%%%%%%%%%
Before proceeding the numerical results, we first briefly review the predictions on mass of $X_b$.
The existence
of the $X_b$ is predicted in both the tetraquark model~\cite{Ali:2009pi} and those involving a molecular interpretation~\cite{Tornqvist:1993ng,Guo:2013sya,Karliner:2013dqa}.
In Ref.~\cite{Ali:2009pi}, the mass of the lowest-lying $1^{++}$ $\bar b \bar q bq$ tetraquark is predicated to be $10504$ MeV , while the mass of the $B\bar B^*$
molecular state is predicated to be a few tens of MeV
higher~\cite{Tornqvist:1993ng,Guo:2013sya,Karliner:2013dqa}. For example, in Ref.~\cite{Tornqvist:1993ng},  the
mass was predicted to be $10562$~MeV, which corresponds to a binding energy to be $42$ MeV, while the
mass was predicted to be $(10580^{+9}_{-8})$~MeV, which corresponds to a binding
energy $(24^{+8}_{-9})$ MeV in Ref.~\cite{Guo:2013sya}.  As can be seen from the theoretical predictions, it might be a good approximation and might be applicable if the binding energy is less than $50$ MeV. In order to cover the range the previous molecular and tetraquark predictions on Ref.~\cite{Ali:2009pi,Tornqvist:1993ng,Guo:2013sya,Karliner:2013dqa}, we present our results up to a binding energy of $100$ MeV, and we will choose several illustrative values: $\epsilon_{X_b} = (5,10,25,50,100)$ MeV.

In Table~\ref{tab:results-5S}, we list the predicted branching ratios by
choosing the monopole and dipole form factors and three values for the
cutoff parameter in the form factor. As a comparison, we also list the predicted branching ratios in NREFT approach. From this table, we can see that the branching ratios for $\Upsilon(5S) \to \gamma X_b$ are orders of $10^{-5}$. The results are not sensitive to both the form factors and the cutoff parameter we choose.

In Fig.~\ref{fig:BrOnEXb} (a), we plot the the branching ratios for $\Upsilon(5S) \to \gamma X_b$ in terms of the binding energy $\epsilon_{X_b}$ with the monopole form factors $\alpha=2.0$ (solid line), $2.5$ (dashed line), and $3.0$ (dotted line), respectively. The coupling constant of $X_b$ in Eq.~(\ref{eq:coupling-Xb}) and the threshold effects can simultaneously influence the binding energy dependence of the branching ratios. With the increasing of the binding energy $\epsilon_{X_b}$,  the coupling strength of $X_b$ increases, and the threshold effects decrease.  Both the coupling strength of $X_b$ and the threshold effects vary quickly in the small $\epsilon_{X_b}$ region and slowly in the large $\epsilon_{X_b}$ region. As a result, the behavior of the branching ratios is relatively sensitive at small $\epsilon_{X_b}$, while it becomes smooth at large $\epsilon_{X_b}$. Results with the dipole form factors $\alpha=2.0$, $2.5$, and $3.0$ are shown in Fig.~\ref{fig:BrOnEXb} (b) as solid, dash, and dotted curves, respectively. The behavior is similar to that of Fig.~\ref{fig:BrOnEXb} (a).

We also predict the branching ratios of $\Upsilon(6S) \to \gamma X_b$ and present the relevant numerical results in Table~\ref{tab:results-6S} and Fig.~\ref{fig:BrOnEXb_6S} with the monopole and dipole form factors.  At the same cutoff parameter $\alpha$, the predicted rates for $\Upsilon(6S)\to \gamma X_b$ are a factor of 2-3  smaller than the corresponding rates for $\Upsilon(5S) \to \gamma X_b$. It indicates that the intermediate $B$-meson loop contribution to the process $\Upsilon(6S)\to \gamma X_b$ is smaller than that to $\Upsilon(5S) \to \gamma X_b$. This is understandable since the mass of $\Upsilon(6S)$ is more far away from the thresholds
of $B^{(*)}B^{(*)}$ than the $\Upsilon(5S)$. But their branching ratios are also about orders of $10^{-5}$ with a reasonable cutoff parameter $\alpha=2\sim 3$.

In Ref.~\cite{Guo:2009wr}, authors introduced a nonrelativistic effective field theory
method to study the meson loop effects
of $\psi^\prime \to J/\psi \pi^0$. Meanwhile they proposed a power counting scheme
to estimate the contribution of the loop
effects, which is used to judge the impact of the
coupled-channel effects. For the diagrams in Fig.~\ref{fig:loops}, the vertex involving the initial bottomonium is in $P$-wave. The momentum in this vertex is contracted with the final photon momentum $q$, and thus should be counted as $q$. The decay amplitude scales as follows,
\begin{eqnarray}
\frac {v^5} {(v^2)^3} q^2 \sim \frac {q^2} {v},
\end{eqnarray}
where $v$ is understood as the average velocity of the
intermediate bottomed mesons.

As a cross-check, we also present the branching ratios
of the decays in the framework of NREFT. The relevant
transition amplitudes are similar to that given in Ref.~\cite{Guo:2013zbw} with only different masses and coupling constants. The obtained numerical results for $\Upsilon(5S) \to \gamma X_b$ and $\Upsilon(6S) \to \gamma X_b$ in terms of the binding energy are listed in the last column of Table \ref{tab:results-5S} and \ref{tab:results-6S}, respectively. As shown in Table \ref{tab:results-5S}, except for the largest binding energy $\epsilon_{X_b}=100$ MeV, the NREFT predictions of $\Upsilon(5S) \to \gamma X_b$
are about $1$ order of magnitude smaller than the ELA results at the commonly accepted range. For $\Upsilon(6S) \to \gamma X_b$ shown in Table \ref{tab:results-6S}, the NREFT predictions are several times smaller than the ELA results in small binding energy range, while the predictions of these two methods are comparable at large binding energy. These difference may give  some sense of the theoretical uncertainties for the predicted rates and indicates the viability of our model to some extent.

Here we should notice, for the isoscalar $X_b$, the pion exchanges might be nonperturbative and produce sizeable effects~\cite{Guo:2013sya,Mehen:2011yh,Valderrama:2012jv}. In Ref.~\cite{Guo:2013sya}, their calculations show that the relative errors of $C_{0X}$ are about 20\% for the $X_b$.  Even if we take into account this effect, the estimated order of the magnitude for the branching ratio $\Upsilon(5S,6S) \to \gamma X_b$ may also be sizeable, which may be measured in the forthcoming BelleII experiments.
%%%%%%%%%%%%%%%%%%
\section{Summary}
\label{sec:summary}
%%%%%%%%%%%%%%%%%%
In this work, we have investigated the production of $X_b$ in the radiative decays of $\Upsilon(5S,6S)$. Based on the  $B {\bar B}^*$ molecular state picture, we considered its production through the mechanism with intermediate
bottom meson loops. Our results have shown that the production ratios for the $\Upsilon(5S,6S) \to \gamma X_b$ are about orders of $10^{-5}$ with a commonly accepted cutoff range $\alpha=2\sim 3$. As a cross-check, we also calculated the branching ratios
of the decays in the framework of NREFT. Except for the large binding energy, the NREFT predictions of $\Upsilon(5S) \to \gamma X_b$
are about $1$ orders of magnitude smaller than the ELA results. The NREFT predictions of $\Upsilon(6S) \to \gamma X_b$ are several times smaller than the ELA results in small binding energy range, while the predictions of these  two methods are comparable at large binding energy. In Ref.~\cite{Li:2014uia,Li:2015uwa}, we have studied the radiative decays and the hidden bottomonium decays of $X_b$. If we consider that the branching ratios of the isospin conserving process $X_b \to \omega \Upsilon(1S)$ are relatively large,  a search for $\Upsilon(5S)\to \gamma X_b\to \gamma \omega \Upsilon(1S)$ may be possible for the updated BelleII experiments. These studies may help us
investigate the $X_b$ deeply. The experimental observation of $X_b$ will provide us with further
insight into the spectroscopy of exotic states and is helpful to probe the structure of the states connected by
the heavy quark symmetry.

%%%%%%%%%%%%%%%%%%
\section*{Acknowledgements}
\label{sec:acknowledgements}
%%%%%%%%%%%%%%%%%%

This work is supported in part by the National Natural Science Foundation of China (Grant Nos. 11275113, 11575100, 11505104)  and the Natural Science Foundation of
Shandong Province (Grant No. ZR2015JL001).

\end{document}